\documentclass[a4paper,11pt]{article}
\pdfoutput=1 % if your are submitting a pdflatex (i.e. if you have
\usepackage{jinstpub} % for details on the use of the package, please
\usepackage{lineno} %\linenumbers

\title{\boldmath GEM Detectors for the CMS Endcap Muon System: \\ status of three new detector stations}

\author{Piet Verwilligen}
\affiliation{INFN sezione di Bari,\\Via E.Orabona 4, I-70125 Bari, Italy}

\emailAdd{piet.verwilligen@ba.infn.it}

\abstract{
The High-Luminosity LHC (HL-LHC, or \textit{Phase-2} LHC) will deliver proton-proton collisions at 5-7.5 times the nominal LHC luminosity, with an expected number of 140-200 pp-interactions per bunch crossing (Pile-up or PU). To maintain the performance of muon triggering and reconstruction under high background, the forward part of the Muon Spectrometer of the CMS experiment will be upgraded with Gas Electron Multipliers (GEM) and improved Resistive Plate Chambers (\textit{i}RPC) detectors~\cite{CMS-TDR-GE11, CMS-TDR-Mu-Upgr}. A first GEM station (GE1/1) was installed during Long Shutdown 2 (LS2, 2019-2021), a 2$^{\text{nd}}$ station (GE2/1) of Triple-GEM detectors will be installed in winter 2023-24 and 2024-25, while a new 6-layer station (ME0) will be installed in the third Long Shutdown (LS3, 2026-2028). GE11 is considered an \textit{early Phase-2} upgrade as it will reduce the $p_{T}$ threshold by combining GEM and Cathode Strip Chamber (CSC) hits in the forward muon system at twice the LHC design luminosity ($\mathcal{L} = 2 \cdot 10^{34}$\,cm$^{-2}$s$^{-1}$, 50 PU). After a successful start of Run-3 in 2022, with almost 40\,fb$^{-1}$ collected, the commissioning of the GE1/1 detector is nearly complete. Most chambers are operated stably with an efficiency in excess of 95\%, next being the demonstration of the combined CSC-GEM trigger in 2023. The lessons learnt with the first large-area GEM station have lead to improvements in detector and electronics design for the \textit{Phase 2} detectors GE2/1 and ME0. This proceeding will discuss the progress made since last MPGD Conference (MPGD 2019)~\cite{MPGD-2019-Proceedings}, discussing the commissioning and early performance of GE1/1; the design improvements and start of construction of GE2/1; and the R\&D currently ongoing for ME0.
}

\arxivnumber{2303.17244} % only if you have one

\collaboration[c]{on behalf of CMS Muon group}

\proceeding{The 7$^{\text{th}}$ International Conference on Micro Pattern Gaseous Detectors, MPGD2022\\
  December 12-16\\
  Weizmann Institute of Science, Rehovot, Israel}

\notoc

\begin{document}
\maketitle
\flushbottom

% \section{GEM detectors for the CMS Endcap Muon System}
% \label{sec:intro} - can I skip because information given in abstract?

\section{Installation, Commissioning, Operation \& Performance of GE1/1}
\label{sec:ge11}
\par
\begin{figure}[htbp]
\centering % \begin{center}/\end{center} takes some additional vertical space
% \includegraphics[width=.3\textwidth,trim=30 110 0 0,clip]{figures}
% \qquad
% \includegraphics[width=.3\textwidth,origin=c,angle=180]{figures/}
\includegraphics[height=.205\textheight]{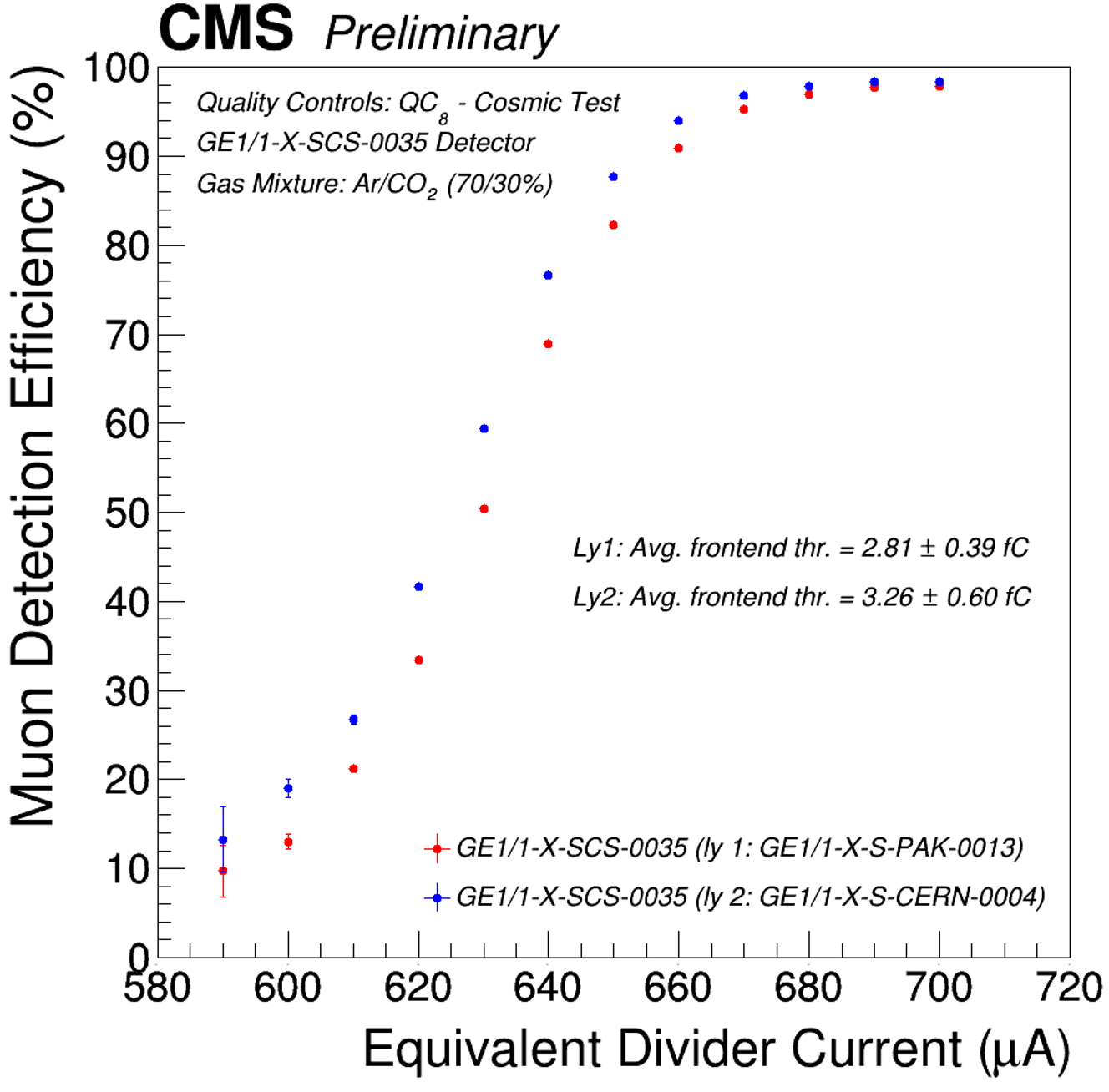}
\includegraphics[height=.205\textheight]{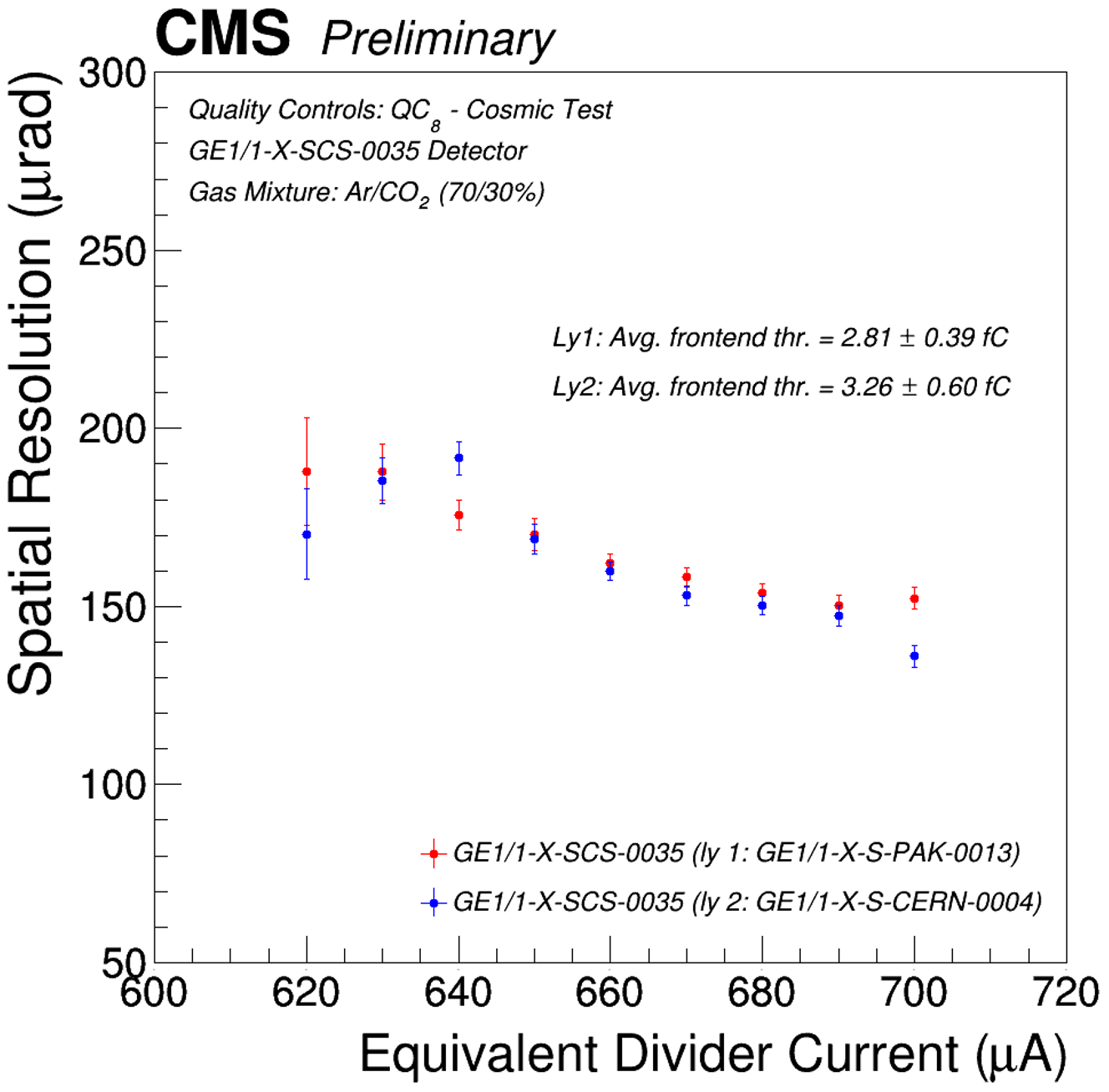}
\includegraphics[height=.195\textheight]{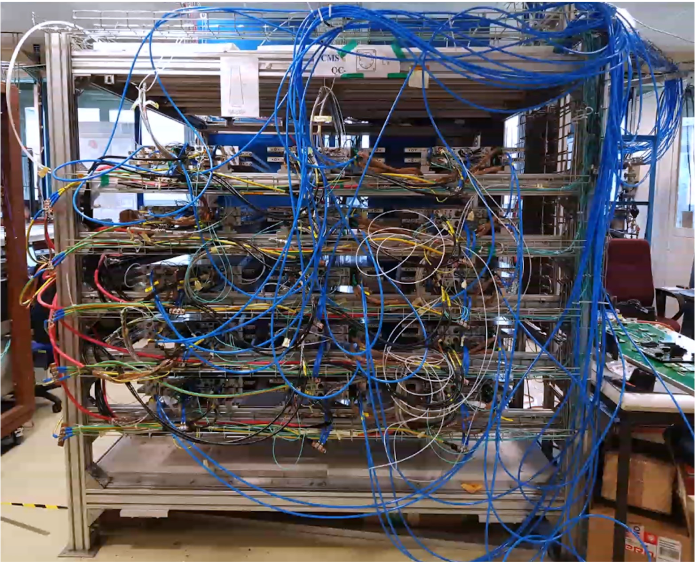}
% "\includegraphics" from the "graphicx" permits to crop (trim+clip)
% and rotate (angle) and image (and much more)
\caption{\label{fig:CosmicStand-Eff} Efficiency (left) and spatial resolution (middle) as function of HV, for two GE1/1 chambers coupled together in a superchamber tested in the Cosmic stand (right). Gain is 20,000 at 690\,$\mu$A, Thr $\approx 3$\,fC.}
\end{figure}

 In 2017-2019 144 (+17 spare) GE1/1 chambers were built and extensively validated: the gain was measured and chambers with near-equal gain were paired in so-called \textit{superchambers}~\cite{NIM-A-2022-166716}. Before installation in CMS, the detectors were installed in the Cosmic Stand (see Fig.~\ref{fig:CosmicStand-Eff}) - a full-fledged mini-experiment with 92k readout channels ($\mu$TCA-based), a 5\,m$^2$ scintillator trigger, with services (HV, LV, gas, cooling) controlled and monitored by a Detector Control System (DCS), running CMS DAQ and Data Quality Monitoring (DQM) - that could test up to 15 superchambers at the time. A HV-scan was performed for all chambers, with about 4M muons per data point (12h run). Figure~\ref{fig:CosmicStand-Eff} shows the efficiency (left) and spatial resolution (right) as function of HV for a selected superchamber. The mean of the efficiency distribution was $> 96$\%. 

\par In 2019-2020 72 superchambers were installed in CMS, at a rate of two superchambers per day. Before and after installation the integrity of the GEM foils was verified applying HV, and a fast electronics test was performed. In the following months the chambers were commissioned without HV, powering electronics and establishing communication. In total 8 superchambers had to be replaced. The GEM foils were trained first in CO$_2$, followed by Ar:CO$_2$ (70:30). In September 2020 the GEMs participated for the first time in Mid-Week Global Runs (MWGR) with cosmic data taking. In Summer 2021 during the Cosmic Run at Zero Tesla (CRuZeT) GEMs took data for $> 250$\,h demonstrating smooth operation and enough data were collected for first alignment and efficiency measurements. In Fall 2021 the GEMs were operated for the first time in 3.8\,T magnetic field during the Cosmic Run at Full Tesla (CRAFT). Several chambers were found to have high discharge rate, which was traced back to metallic dust particles (originating from construction) moved by the strong magnetic field. 2021 commissioning was completed with beam splash and pilot collisions at $\sqrt{s} = 900$\,GeV.

\par
\begin{figure}[htbp]
\centering 
\includegraphics[height=.195\textheight]{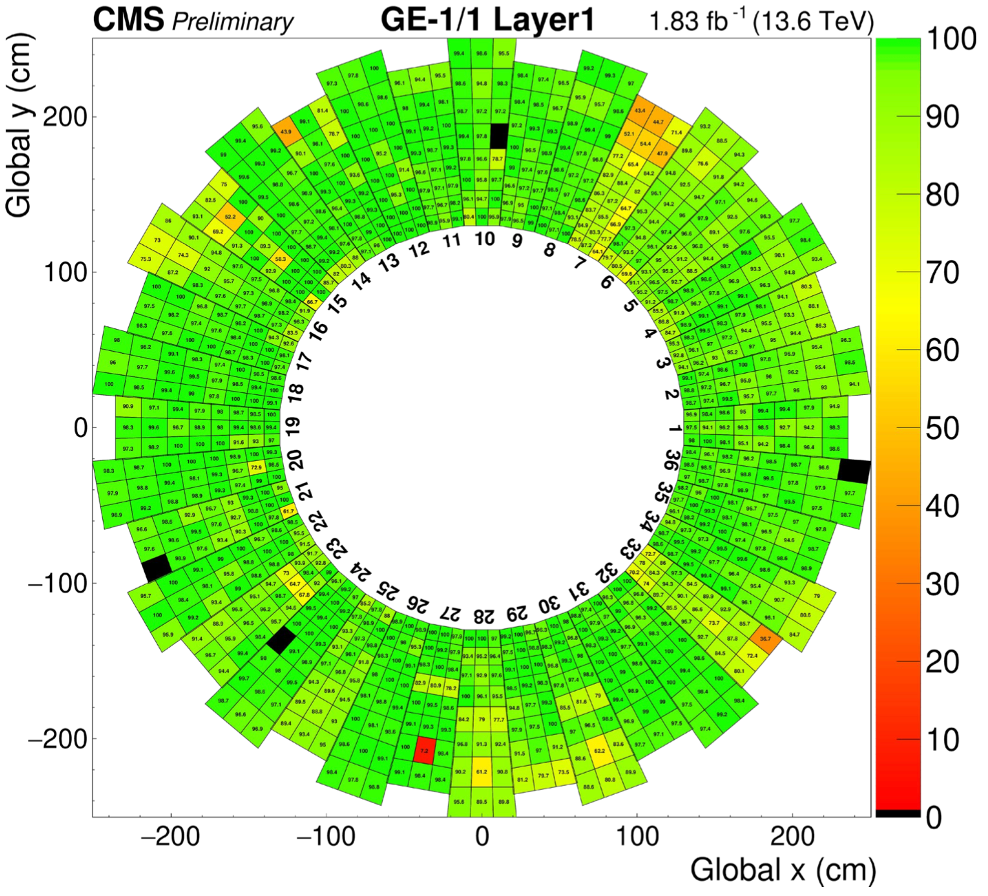}
\includegraphics[height=.205\textheight]{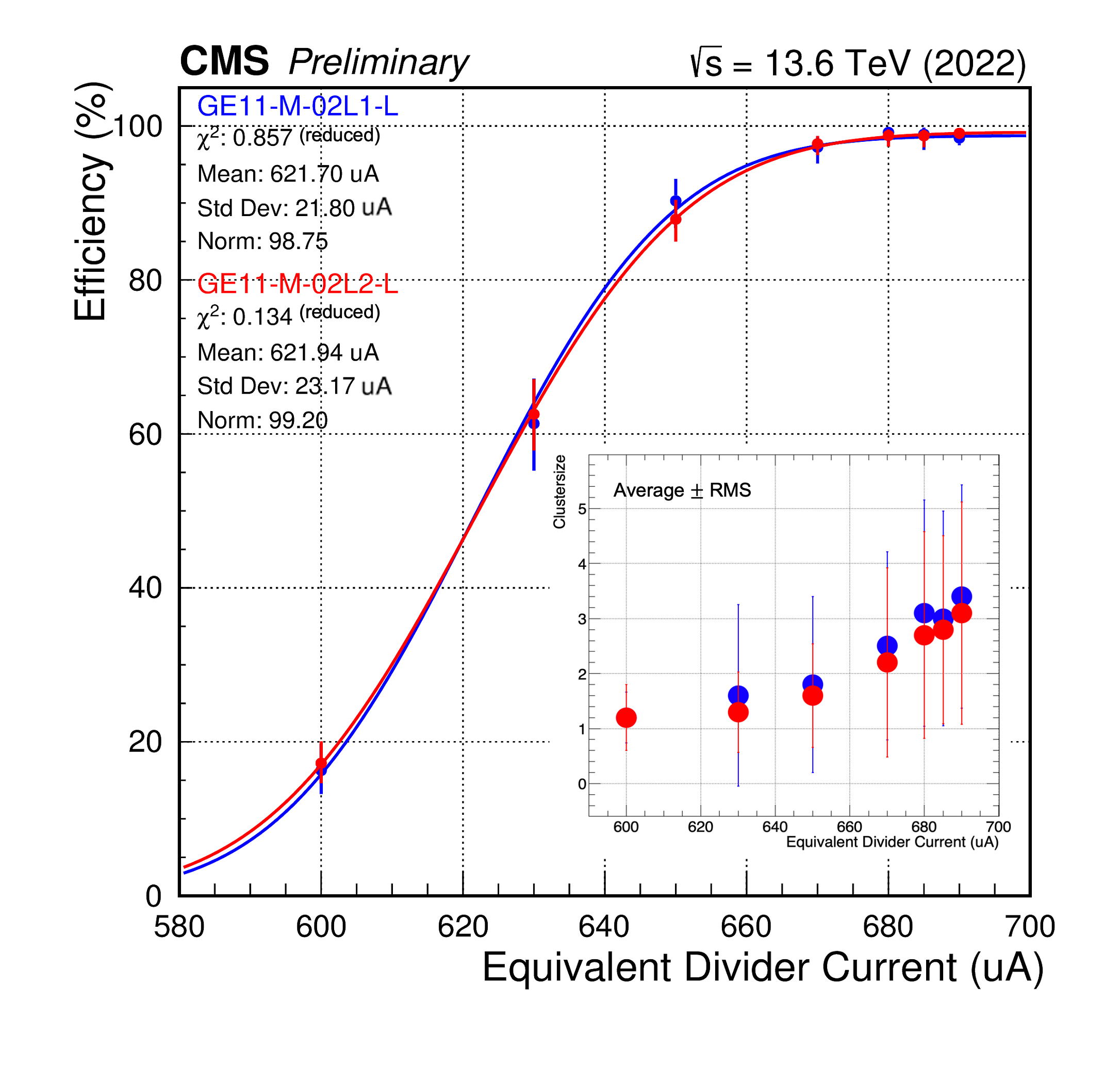}
\includegraphics[height=.205\textheight]{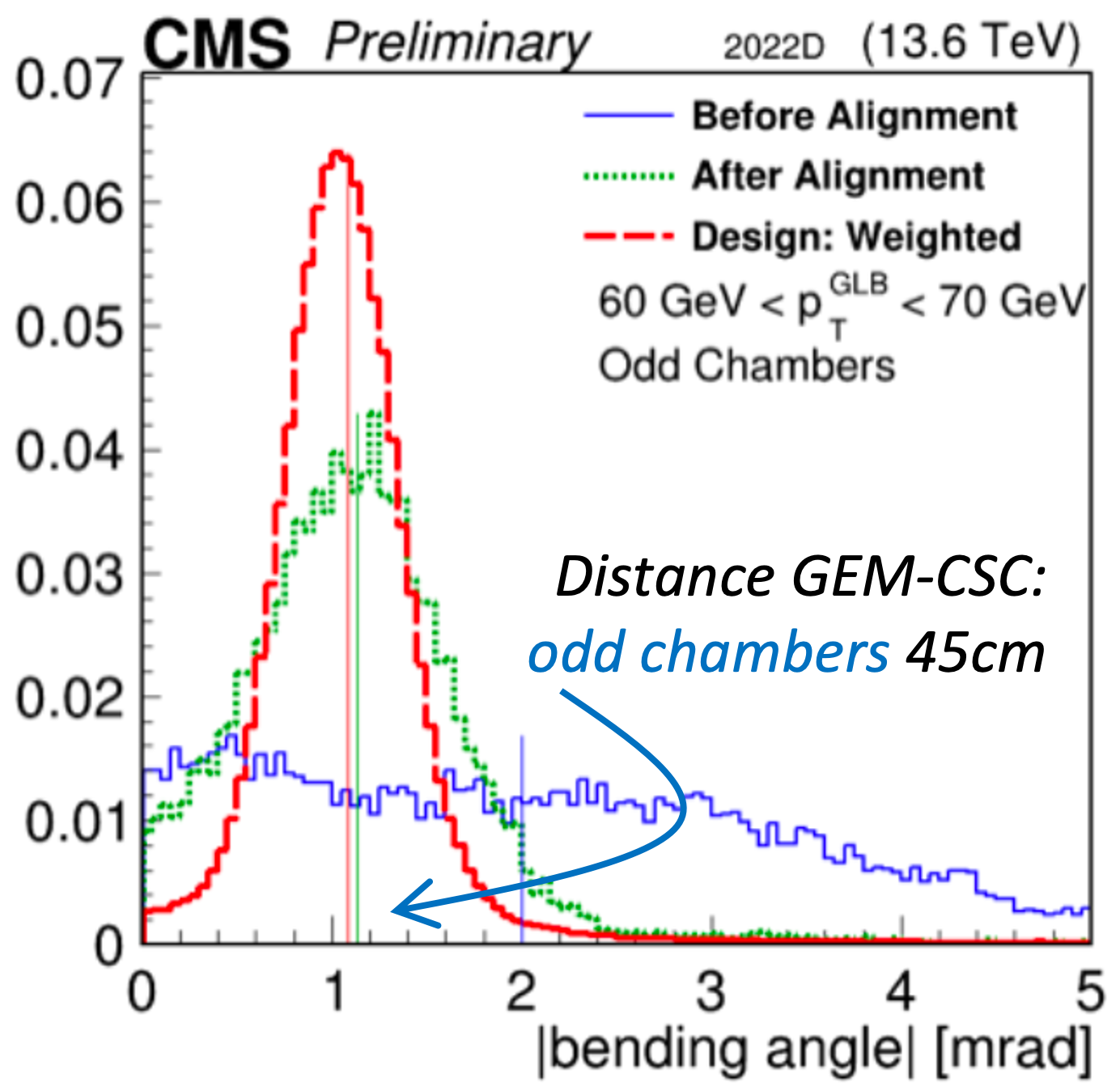}
\caption{\label{fig:HVScan} Efficiency of 36 GE1/1 chambers (negative endcap, first layer) measured with collision muons (left), Efficiency and clustersize for two GE1/1 in a superchamber as function of HV (middle) and Bending Angle for odd chambers measured for muons with $60 < p_\mathrm{T} < 70$\,GeV/c (right).}
\end{figure}

 July 12$^{\text{th}}$ 2022 marked the start of Run-3 with more than 1000 colliding bunches (\textit{cb}) in the LHC. The instantaneous luminosity was fastly ramped from $3.5 \times 10^{32}$\,cm$^{-2}$s$^{-1}$ (75\textit{cb}) to regular running at $1.8 \times 10^{34}$\,cm$^{-2}$s$^{-1}$ (2400\textit{cb}). During this period the background in the GE1/1 chambers increased, leading to an increased number of discharges, which were overcome with a HV-conditioning period. In total 34 sectors were lost due to a short (0.5\%) of the system. A calibration procedure was implemented to check the health of the front-end channels and the number of damaged/lost channels was measured to be $<0.2$\,\%. In Fall 2022 the efficiency of the GE1/1 chambers was measured using muons with $p_T > 20$\,GeV/c from $Z$-decays, and a HV-scan was performed. Figure~\ref{fig:HVScan} shows the efficiency and clustersize for a selected chamber. Enough data were collected to perform a precise alignment and the bending angle between the GEM hits and the CSC segment was measured, paving the way for the combined GEM-CSC trigger in 2023.  

\section{Status of \textit{Phase-2} detectors: GE2/1 \& ME0}
\label{sec:ge21-me0}
\par
\begin{figure}[htbp]
\centering
\includegraphics[height=.200\textheight]{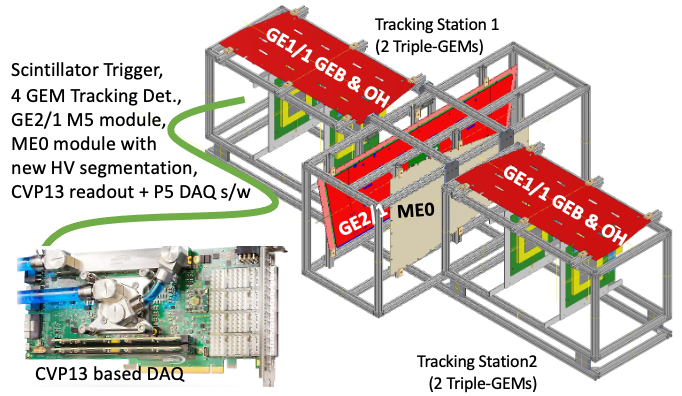}
\includegraphics[height=.205\textheight]{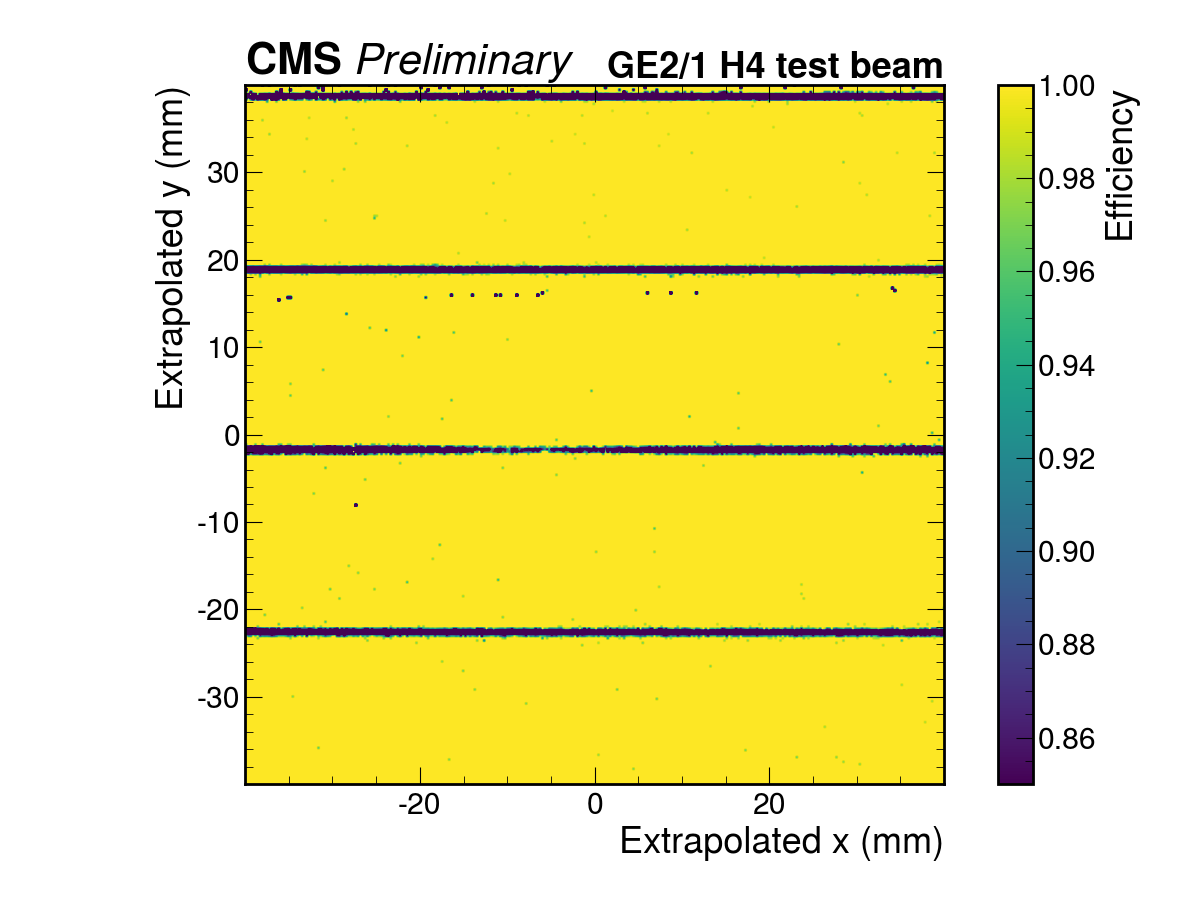}
\caption{\label{fig:Testbeam} Left: Testbeam setup with 4 Triple-GEM trackers (250\,$\mu$m strip pitch - $\sigma = 80$\,$\mu$m) and GE2/1 and ME0 modules under test. Data is sent through optical fibers to a commercial FPGA board (CVP13). Right: 2D Efficiency plot of GE2/1 under test, dark blue lines are due to dead area between two HV segments.}
\end{figure}

 For each endcap the GE2/1 station will consist of two layers of 18 chambers covering $\Delta \phi = 20^\circ$ and $1.6 < \eta < 2.4$. The chambers are about 1.9\,m long and 0.5-1.3\,m wide. The chambers will consist of 4 detector modules (with max size 0.5\,m $\times$ 1.3\,m) and 1536 readout channels each. The GE2/1 station will improve the measurement of the bending angle in the CSCs of the 2nd muon station and can be used for triggering on displaced muons \cite{CMS-TDR-Mu-Upgr}. Several improvements were made w.r.t. the design of GE1/1: the VFAT3 asics are now packaged chips instead of wirebonded, and are mounted on PCBs with flex connector; the grounding in readout board and electronics board has been improved; each detector module has its own Opto-Hybrid board; the VTRx~\cite{VTRX} is now cooled to prevent outgassing problems; to prevent discharge propagation 1$^\textrm{st}$ and 2$^\textrm{nd}$ GEM foils are now double segmented (100\,k$\Omega$ resistor at bottom, 10\,M$\Omega$ at top) while the 3$\textrm{rd}$ foil is single segmented to reduce cross talk. A first endcap is expected to be ready for installation in winter 2023, currently 51 modules out of 144 have been constructed and tested. A production GE2/1 chamber dressed with final electronics has been tested with beams at SPS in October '21 and May '22 and an efficiency $>97$\% has been measured, while noise levels are $<0.5$\,fC. The setup of the testbeam is shown in Figure~\ref{fig:Testbeam} (left) together with a 2D efficiency plot of a 9\,cm $\times$ 9\,cm window of the GE2/1 chamber under test. More details can be found in Ref.~\cite{Pisa-2021-Testbeam}.

% \section{Research \& Development for ME0}
% \label{sec:me0}
\par The ME0 station is the most challenging upgrade of the muon system with Triple-GEM technology. It will extend the coverage of the muon system up to $\eta < 2.8$, will be installed directly behind the new high-granularity calorimeter, and will consist of a stack of 6 chambers. In total 18 stacks covering $20^\circ$ will be installed in each endcap. These chambers have to work in a rate environment with a large gradient starting at 150\,kHz/cm$^2$ at $\eta = 2.8$  down to 2\,kHz/cm$^2$ at $\eta = 2.0$. Rate capability tests have pointed out that at these high rates gain loss was observed due to the large currents passing through the foil protection resistors and HV filter resistors. The HV segment protection resistors were lowered to 2\,M$\Omega$, which was found an optimum between a lower voltage drop due to a fixed current on one hand, and not increasing too much the current in case of a short in the HV-segment.

\begin{figure}[htbp]
\centering
\includegraphics[height=.205\textheight]{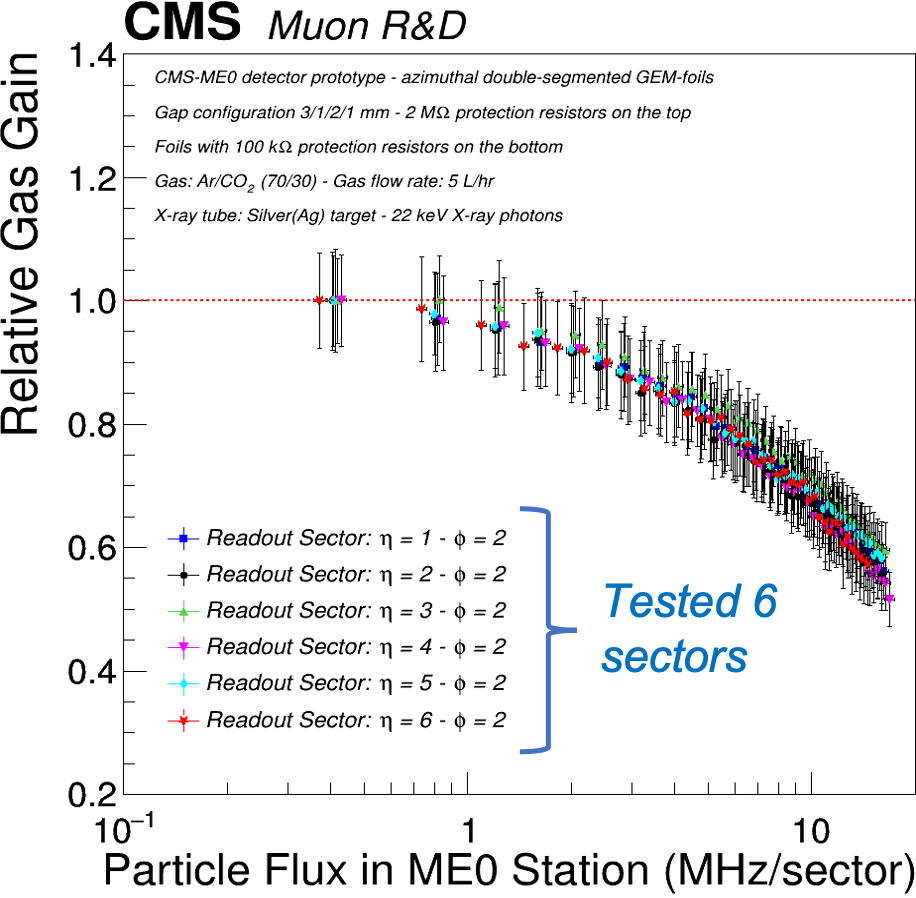}
\includegraphics[height=.205\textheight]{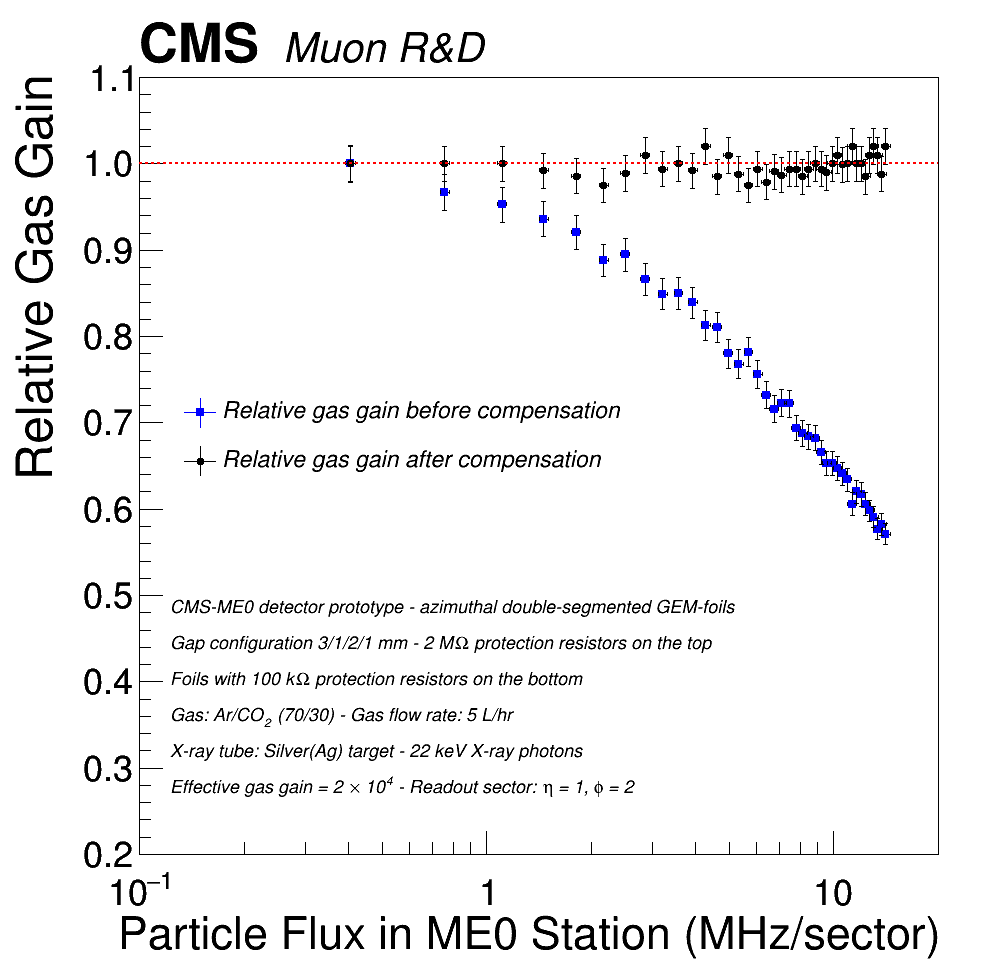}
\includegraphics[height=.205\textheight]{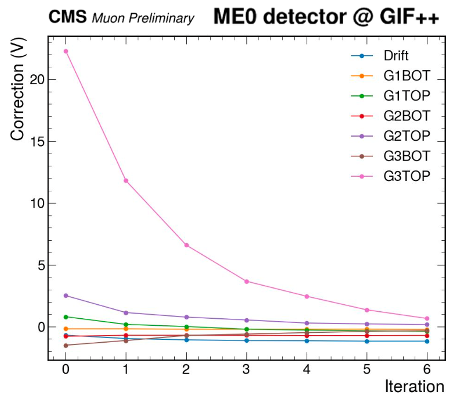}
\caption{\label{fig:GainComp} Left: Gain loss observed with very intense irradiation with X-rays at the bottom of the detector: all sectors - also far away from the source - show equal gain loss. Middle: Compensation of the gain by increasing HV. Right: Iterative procedure for the $\Delta$HV to be applied to compensate the gain.}
\end{figure}

\par To equalize the currents, the GEM foil was divided in 40 azimuthal segments, and in simulation was found that the expected particle rate per HV segment is between 1.6 and 1.7\,MHz. This allows to compensate for gain loss due to the voltage drop over the protection resistors by incrementing the HV in an iterative procedure over all GEM electrodes (see Figure~\ref{fig:GainComp}). Three prototype ME0 chambers with azimuthal segmentation were build and the gain restauration principle was demonstrated in tests with X-rays and with beam under irradiation at the GIF++ in July '22.

\par The change in segmentation of the GEM foils from horizontal (GE1/1, GE2/1) to azimuthal (ME0) has lead to an increase of dead area for two reasons: (1) the segments run now along the long side of the detector; (2) the long edges of the trapezoidal segments are now not horizontal anymore, and this leads to an irregular pattern of GEM holes near the border, leading sometimes up to $\sim600$\,$\mu$m of distance between the closest GEM holes of two neighbouring segments. The overall efficiency of an ME0 with this so-called \textit{blank} segmentation is $\sim96$\%. To reduce this effect GEM foils with so-called \textit{random-hole} segmentation were produced and tested. In this technique the hole pattern is continuous over the entire GEM foil and in the zones between two HV segments only the copper layer was removed. More details can be found in Ref.\cite{MPGD-2022-Random} in the same proceedings. The efficiency of the ME0 prototype with \textit{random-hole} segmentation was found to be 97.5\%.

\par The rate capability of the ME0 prototype with azimuthal segmentation was verified under intense radiation at the GIF++. At the highest rates to be expected in CMS (150\,kHz/cm$^2$ - in correspondence with the area of the detector closest to the beampipe) a few percent of efficiency loss was observed, independent of the gain compensation. This loss was found in agreement with the $\sim400$\,ns deadtime of the VFAT3 front-end electronics. As this affects only a very small fraction of the chamber no further action was taken. As a ME0 stack consists of 6 layers, and a 4-out-of-6 algorithm will be implemented for segment finding, the effect of the electronics dead time will be minimized further.

% \begin{figure}[htbp]
% \centering
% \includegraphics[height=.200\textheight]{figures/Fig-4-B-ME0HVSeg.png}
% \includegraphics[height=.205\textheight]{figures/Fig-4-A-ME0-TB.png}
% \includegraphics[height=.205\textheight]{figures/Fig-4-C-ME0-Rand-TB.jpg}
% \includegraphics[height=.205\textheight]{figures/Fig-4-A-ME0-TB.png}
% \caption{\label{fig:ME0TB} caption.}
% \end{figure}
  
\acknowledgments
The author would like to thank the RD-51 collaboration for having started this great conference series on MPGDs and to the RD-51 colleagues of the Weizmann Institute of Science for the excellent organization of MPGD2022, for the friendly atmosphere, good food and great social events. Special thanks go to all his colleagues working on the GEM upgrade of the CMS muon system and colleagues in RD-51. The CMS GEM group gratefully acknowledges support from FRS-FNRS (Belgium), FWO-Flanders (Belgium), BSF-MES (Bulgaria), MOST and NSFC (China), BMBF (Germany), DAE (India), DST (India), INFN (Italy), NRF (Korea), QNRF (Qatar), and DOE (USA). 
% Finally I would like to thank all colleagues in RD-51 for the inspiring discussions and help whenever needed.

% \paragraph{Note added.} This is also a good position for notes added after the paper has been written.



\begin{thebibliography}{99}

\bibitem{CMS-TDR-GE11}
Colaleo, A \emph{et al.}, \emph{CMS Technical Design Report for the Muon Endcap GEM Upgrade}, \emph{CERN-LHCC-2015-012, CMS-TDR-013} (2015) \url{https://cds.cern.ch/record/2021453}.

  
\bibitem{CMS-TDR-Mu-Upgr}
CMS Collaboration, \emph{The Phase-2 Upgrade of the CMS Muon Detectors}, \emph{CERN-LHCC-2017-012, CMS-TDR-016} (2017), \url{https://cds.cern.ch/record/2283189}.
  
\bibitem{MPGD-2019-Proceedings}
6$^{\text{th}}$ International Conference on Micro Pattern Gaseous Detectors (MPGD2019), 5-10 May 2019, La Rochelle, France. Proceedings published in \emph{J.Phys.Conf.Ser. 1498 (2020)}, \url{https://indico.cern.ch/event/757322/}. %\url{https://inspirehep.net/conferences/1704046}.

\bibitem{NIM-A-2022-166716}
  M.Abbas \emph{et al.}, \emph{Quality control of mass-produced GEM detectors for the CMS GE1/1 muon upgrade}, \emph{Nucl.Instrum.Meth.A} 1022 (2022) 166716, \texttt{doi:10.1016/j.nima.2022.166716}.

\bibitem{VTRX}
  C.Soos \emph{et al.}, \emph{Versatile Link Plus transceiver development}, \emph{JINST} 12 C03068, \texttt{doi:10.1088/1748-0221/12/03/c030668}.

\bibitem{Pisa-2021-Testbeam}
Pellecchia, A, \emph{et al.}, \emph{Performance of triple-GEM detectors for the CMS Phase-2 upgrade measured in test beam}, \emph{Nucl.Instrum.Meth.A} 1046 (2023) 167618, \texttt{doi:10.1016/j.nima.2022.167618}.

\bibitem{MPGD-2022-Random}
Pellecchia, A \emph{et al.}, \emph{Production and characterization of random electrode sectorization in GEM foils}, submitted to JINST (same proceedings), \texttt{arXiv:2303.06355}.
% \bibitem{a}
% Author, \emph{Title}, \emph{J. Abbrev.} {\bf vol} (year) pg.

% \bibitem{b}
% Author, \emph{Title},
% arxiv:1234.5678.

% \bibitem{c}
% Author, \emph{Title},
% Publisher (year).


% Please avoid comments such as "For a review'', "For some examples",
% "and references therein" or move them in the text. In general,
% please leave only references in the bibliography and move all
% accessory text in footnotes.

% Also, please have only one work for each \bibitem.


\end{thebibliography}
\end{document}